\documentclass[10pt, conference, letterpaper]{IEEEtran}
\IEEEoverridecommandlockouts
\usepackage{algpseudocode}
\usepackage[linesnumbered,ruled]{algorithm2e}
\usepackage{algorithmicx}

\usepackage{stfloats}
\usepackage[caption=false,font=footnotesize]{subfig}

\usepackage{float}
\usepackage{cite}
\usepackage{amsmath,amssymb,amsfonts}
\usepackage{graphicx}
\usepackage{textcomp}
\usepackage{xcolor}
\def\BibTeX{{\rm B\kern-.05em{\sc i\kern-.025em b}\kern-.08em
    T\kern-.1667em\lower.7ex\hbox{E}\kern-.125emX}}
    
\IEEEoverridecommandlockouts

\begin{document}

\title{Index Coding at the WiFi Edge: An Implementation Study for Video Delivery\\
\thanks{This work was supported in part by a SERB grant on “Content Caching and Delivery over Wireless Networks”.}
}

\author
{
\IEEEauthorblockN{Lalhruaizela Chhangte% \\ Dept. of EE
}
\IEEEauthorblockA{IITB-Monash Research Academy \\ Mumbai, India \\ 154074004@iitb.ac.in}
\and
\IEEEauthorblockN{Emanuele Viterbo %\\ Dept. of ECSE
}
\IEEEauthorblockA{Monash University \\ Clayton, VIC, Australia \\ emanuele.viterbo@monash.edu}
\and
\IEEEauthorblockN{D Manjunath%\\ Dept. of EE
}
\IEEEauthorblockA{IIT Bombay \\ Mumbai, India \\ dmanju@ee.iitb.ac.in}
\and
\IEEEauthorblockN{Nikhil Karamchandani% \\ Dept. of EE
}
\IEEEauthorblockA{IIT Bombay \\ Mumbai, India \\ nikhilk@ee.iitb.ac.in}
}
\IEEEaftertitletext{\vspace{-2\baselineskip}}
\maketitle
\begin{abstract}

HTTP based video streaming has become the de facto standard for video content delivery across different video streaming services. However, video content delivery continues to be challenged at the wireless edge by inadequate and highly variable bandwidth. In this paper, we describe WiCode, a platform that improves HTTP based video content delivery at the WiFi edge. WiCode uses coded delivery at the WiFi AP to reduce data transmissions in order to improve the perceived performance of video streaming at the users. WiCode performs index coding on video segments to reduce the number of bits transmitted. Further, it also performs index coding on UDP packets that are retransmitted to reduce the number of bits transmitted. This paper describes the design and implementation of WiCode, and the practical gains achievable due to employing coded delivery in a real system taking into account the overheads introduced by WiCode. The WiCode module at the client side is a browser plugin that does not require any client side device configuration changes. We also show the effect of variable and fixed length segment size on the perceived performance of WiCode.

\end{abstract}

%\begin{IEEEkeywords}
%index coding, edge caching, DASH
%\end{IEEEkeywords}

\section{Introduction}
Video content dominates the global IP traffic, and it is expected to reach 82\% by 2020 \cite{cisco}. Also, the majority of video content today is delivered by video streaming services using HTTP based streaming technologies such as DASH \cite{dash}, Microsoft's Smooth Streaming \cite{microsoft}, Adobe's HTTP Dynamic streaming \cite{adobe}, or Apple's HTTP Live streaming \cite{apple}. In HTTP based streaming, a video file is split into multiple segments and stored at the server, and each segment is requested at an appropriate time, and delivered using HTTP \cite{httpdash}. Currently, there is a significant challenge in delivering video content with satisfactory quality of service at the wireless edge due to inadequate and highly variable bandwidth. This challenge is exacerbated because wireless devices are expected to account for two thirds of the global IP traffic by 2020 \cite{cisco}.

There are several proposals in the literature to reduce the bandwidth usage of the last hop broadcast network. Our interest in this paper is on the information-theoretic approach where the server broadcasts index-coded information simultaneously to multiple clients\cite{iscod,indexside}. This broadcast information is simultaneously received by multiple clients who decode the information that they need by using the side information (previously downloaded information) in their local caches. Coding and decoding of content in coded delivery is achieved by simple XOR operation. Coded delivery has been found to improve content delivery by several studies \cite{fundamental, decentralized, delaysensitive}. However, these studies have remained mostly of theoretical interest, and therefore, to the best of our knowledge, there are no existing systems that uses coded delivery at the WiFi edge to improve HTTP based video content delivery.

In this work, we design and build a coded delivery system for a wireless network, specifically a WiFi network to improve HTTP based video content delivery. Specifically, the following are the contributions in this paper.

\begin{itemize}
\item Design of a platform that uses index coding to deliver coded video segments to groups of users at the WiFi edge.
\item An index coding module as a browser plugin on the client side that does not require device configuration changes.
\item A wide range of measurements that illustrate the effectiveness of the platform, and also the effect of variable and fixed length segment sizes on the performance of the system.
\end{itemize}
The rest of the paper is organized as follows. In the next section we provide a brief overview and design of WiCode. Section~\ref{sec:implement} provides the implementation details of WiCode system, and Section~\ref{sec:experiments} describes the experimental setup and key results.

%, and DASH clients do not use HTTP byte range request for video
%segment requests. Also, the experiments that are carried out is
%restricted to a single quality level for the video segments. This
%work explores the implementation challenges of index coding for DASH,
%the performance achievable in terms of throughput, bits transmitted
%and extra computation required in a real system.
\vspace{-2mm}
\section{WiCode Overview}
\label{sec:overview}

\begin{figure}%[H]
  \centering
  \includegraphics[scale=0.55]{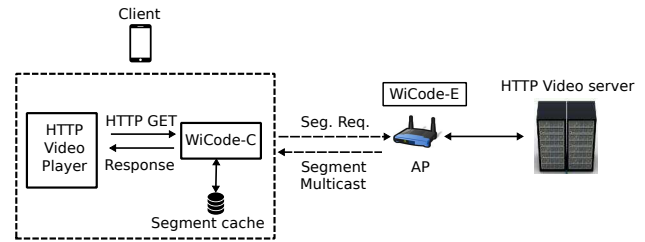}
  \caption{WiCode overview.}
  \vspace{-1mm}
  \label{fig:overview}
  \vspace{-6mm}
\end{figure}

WiCode consists of two modules---WiCode-C, a module at the client, and WiCode-E, a module at the WiFi access point (AP); see Fig.~\ref{fig:overview} for an illustration. A HTTP GET request for a video segment from a video player is intercepted by WiCode-C at the client, which forwards it to WiCode-E at the WiFi AP. WiCode-E then fetches the requested video segment from the HTTP video server, which is eventually transmitted to the client. WiCode-E transmits video segments over the WiFi network using multicast, and WiCode-C at the clients process multicast segments that are only designated for them. The multicast video segment can be either index-coded or non-index-coded. WiCode-C decodes the requested segment from the received index-coded segment by performing XOR with the side information (segment) from the segment cache. The decoded segment is then passed on to the video player as an HTTP response. 

%In the following sections, we describe in detail how WiCode works. First we provide a formal description of index coding in the context of HTTP based video streaming. Second, we describe how asynchronous video segment requests are handled at the AP, and the index coding heuristic algorithm used. Third we explain the effect of variable length segment size on the efficiency of index coding, and finally we describe how segment multicasting and retransmissions are implemented in WiCode. 

\subsection{Index coding model}
\label{subsec:index}
Consider a server that has $N$ video files denoted by $F := \{f_{1}, \ldots, f_{N}\}$ and $M$ clients denoted by $C := \{c_{1}, \ldots, c_{M}\}$. Each video file is split into a number of video segments of fixed playback duration. Streaming a video file $f_{i}$ that is split into $S$ number of video segments involves requesting each of the video segments $f_{i}^{(1)}, f_{i}^{(2)}, \ldots, f_{i}^{(S)}$ one after the other.  A video segment request from client $c_{i}$ can be represented as a pair $(W(c_{i}), H(c_{i}))$, where $W(c_{i})$ is the set of video segments that $c_{i}$ requests, and $H(c_{i})$ is the set of segments that is cached (previously requested) at $c_{i}$. The requests from $c_{i}$ and $c_{j}$ can be index-coded if 
\begin{align*}
  W(c_{i}) \subset H(c_{j}) \mbox{ and } W(c_{j}) \subset H(c_{i}).
\end{align*}
Given $\{W(c_{i}),H(c_{i})\}_{i=1,\ldots,M},$ determining the optimal policy that maximizes index coding is NP-hard \cite{indexalgorithms}. However, there are many low complexity heuristic algorithms that perform well \cite{indexalgorithms}.

In a HTTP based video streaming, $|W(c_{i})|$ is always 1, i.e., only one video segment is requested at a time from a client. Also, $H(c_{i})$, the cache content information at a client has to be sent to the node that performs index coding. In WiCode, this information is sent to WiCode-E by WiCode-C every time a client joins the WiFi network, and WiCode-E keeps track of the cache content as long as the client is associated to the AP.  

\vspace{-1mm}

\subsection{Asynchronous requests and proactive coding}
\label{subsec:asyn}

\begin{figure}
  \centering
  \subfloat[Without proactive coding]{\includegraphics[width=0.20\textwidth]{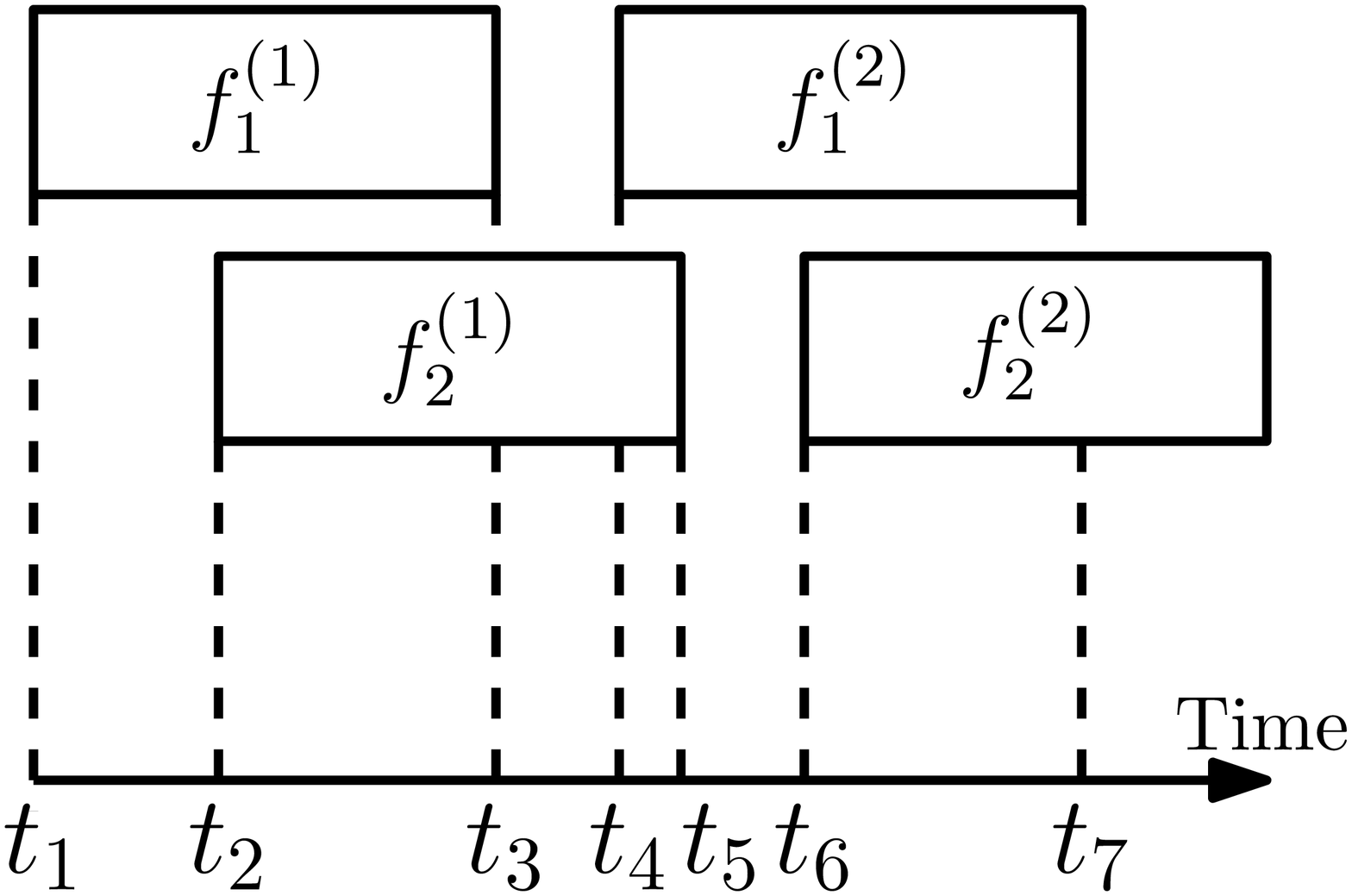}\label{fig:proactive_1}}\qquad
  \subfloat[With proactive coding]{\includegraphics[width=0.20\textwidth]{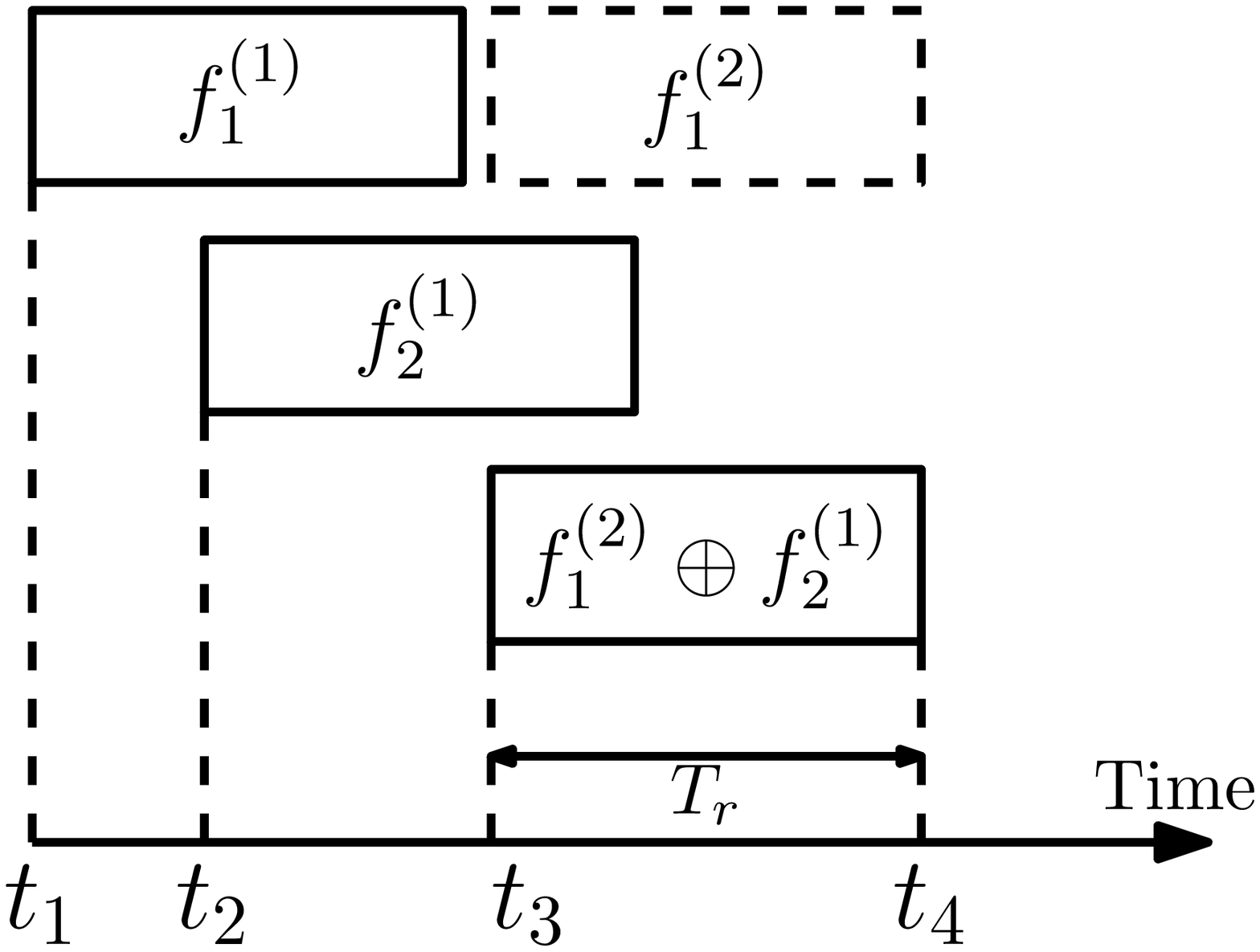}\label{fig:proactive_2}}
  \caption{Asynchronous requests and proactive coding at WiCode-E.}
  \label{fig:proactive}
  \vspace{-5mm}
\end{figure}

\begin{figure}[H]
  \centering
  \includegraphics[scale=0.3]{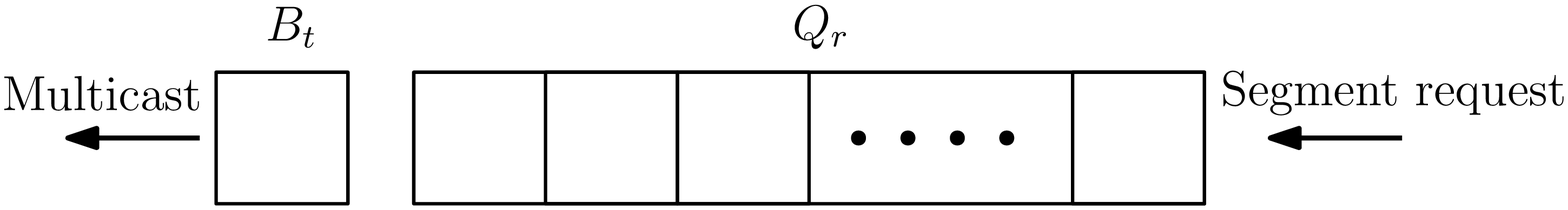}
  \caption{Illustrating $Q_{r}$ and $B_{t}$ at WiCode-E.}
  \label{fig:transmission}
  \vspace{-3mm}
\end{figure}

In a wireless network, the number of clients changes frequently, and different clients request video segments at different times. Therefore, video segment requests arrive at the server asynchronously. To handle the asynchronous arrival of requests, WiCode-E puts the incoming requests from the clients in a queue, $Q_{r}$, where the requests wait for their transmission, and the segment request currently being transmitted is placed in a transmission buffer $B_{t}$ as illustrated in Fig. \ref{fig:transmission}.

%\vspace{-8mm}
For any incoming request $r$, WiCode-E checks whether $r$ can be index coded with any request $s$ currently in $Q_{r}$, where $s$ can be non-index-coded or index-coded (two or more segments coded together). WiCode uses the algorithm illustrated in Algorithm \ref{algo:index}, which uses a greedy approach, and it maximizes the number of segments coded together in a coded segment.

\begin{algorithm}%[t]
  \caption{Index coding}
  {\small
  \label{algo:index}
  \SetAlgoLined
  %\KwResult{Write here the result }
  I: set of segment requests that can be coded with $r$\;
  J: set of segment requests from I with maximum number of segments coded together\;
  Pick the segment $s$ from J that arrives the earliest at WiCode-E and code with $r$\;
  }
 \vspace{-1mm}
\end{algorithm}
\vspace{-3mm}

The asynchronous arrival of requests also results in loss of coding opportunities. To illustrate the issue, consider a simple case of two clients that request two different video streams, $f_{1}$ and $f_{2}$ as shown in Fig. \ref{fig:proactive_1}. Further, assume that these requests can be coded together. The requests though arrive as follows: Request $f_{1}^{(1)}$ arrives at time $t_{1}$ and it will be placed in the transmission buffer $B_{t}$ for transmission. $f_{2}^{(1)}$ arrives at time $t_{2}$ while $f_{1}^{(1)}$ is being transmitted, and therefore they cannot be coded together. Transmission of $f_{1}^{(1)}$ ends at time $t_{3}$ and $f_{2}^{(1)}$ will be placed in $B_{t}$, and the transmission of $f_{2}^{(1)}$ begins. Again, request $f_{1}^{(2)}$ arrives at time $t_{4}$, and again it cannot be coded with $f_{2}^{(1)}$ resulting in loss of coding. The lost in coding opportunities can continue for the rest of the segment requests from these clients. To mitigate this loss, the transmission of $f_{2}^{(1)}$ can be delayed arbitrarily so that it may be coded with future requests, in this case, $f_{1}^{(2)}.$ However, delaying every segment transmission with the expectation that there would be future requests that can be coded with it would also increase the overall latency.

To address the latency-coding-gain tradeoff, WiCode-E uses selective delay in servicing segment requests and some proactive coding. The example illustrated in Fig. \ref{fig:proactive_2} motivates the scheme. Request $f_{1}^{(1)}$ arrives at time $t_{1}$ and its transmission begins immediately. Also, at the same time, if $f_{1}^{(1)}$ is not the last segment in the video stream of $f_{1}$ then WiCode-E proactively places $f_{1}^{(2)}$ in $Q_{r}$ without the actual request coming from the client. If a request for a segment $f_{2}^{(1)}$ arrives before the end of transmission of $f_{1}^{(1)}$ at time $t_{2}$, then $f_{2}^{(1)}$ is coded with $f_{1}^{(2)}$ to form $f_{1}^{(2)} \oplus f_{2}^{(1)}$. If the actual request of $f_{1}^{(2)}$ arrives before a timeout $T_{r}$ after the current transmission ($f_{1}^{(1)}$) is completed, then the coded stream is transmitted, else the uncoded $f_{2}^{(1)}$ is transmitted. If $f_{1}^{(2)} \oplus f_{2}^{(1)}$ is transmitted, then the next segments from both of the video streams will be placed in $Q_{r}$ immediately which ensures index coding of the segments.
\vspace{-2mm}
\subsection{Variable segment size}
\label{subsec:variable}

\begin{figure}%[H]
\vspace{-5mm}
  \centering
  \includegraphics[scale=0.18]{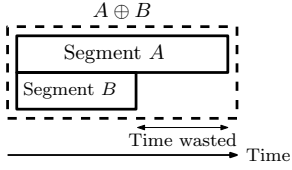}  
  \vspace{-2mm}
  \caption{Illustrating a reduction of gain due to index-coding when the number of bits in the frames being coded together are not equal. Overlapping parts of segments A and B are coded. Total transmission time is the length of segment A with the latter part of A sent as uncoded stream.}
  \label{fig:time_waste}
  \vspace{-5mm}
\end{figure}

In a HTTP based video streaming, video files are split into multiple video segments of fixed playback duration. For example Microsoft Smooth Streaming \cite{microsoft} uses 2 seconds playback duration, while Apple's HTTP Live Streaming \cite{apple} uses 10 seconds. However, the size of video segments that have the same playback duration within the same video file or across different video files can vary widely. The drawback in having variable length segment sizes is that it possibly results in having to code larger segments with smaller segments; this can reduce the achievable coding gain; see Fig.~\ref{fig:time_waste} for an illustration. WiCode addresses this issue by preferring coding of similar sized segments when a choice is available. Even when the segments are of similar size but have different playback duration, we will see that this significantly improves the perceived throughput at the receiver.

%To generate video segments that have a duration $t$, a bitrate $b$ from a video file with size $S_{b}$, and total video duration $D$, a HTTP based video encoder produces $L_{tb}$ video segments, each segment of size

%\begin{displaymath} 
  %S_{tb} = \alpha \frac{t}{D} S_{b}
%\end{displaymath}           
%where $\alpha$ depends on the efficiency of the encoder and the nature of the video frames in the segment. Therefore, if the variation in segment size is very large, it results in index coding of large segments with smaller segments which significantly reduces the encoding efficiency; see Fig.~\ref{time_waste} for an illustration. WiCode addresses the issue by coding a segment with similar size segment when it can be coded with multiple segments.

\subsection{Segment multicasting and retransmission}
\label{subsec:multicast}
\begin{figure}
  \centering
  \includegraphics[width=0.35\textwidth]{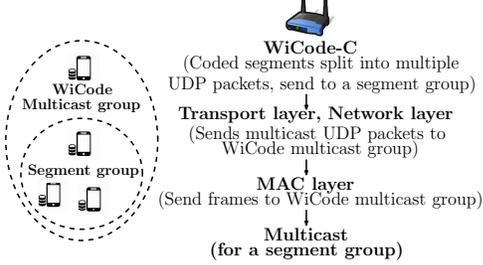}    
  \caption{Illustrating segment multicasting in WiCode.}
  %\vspace{-2mm}
  \label{fig:multicast}
  \vspace{-7mm}
\end{figure}

{\em MAC layer multicasting:} Delivery of coded segment to multiple WiCode clients simultaneously in a WiFi network is achieved by employing multicasting at the MAC layer. However, multicasting or broadcasting in 802.11 network has a key disadvantage---lack of ACK at the MAC layer. This potentially leads to lower effective throughput as 802.11 frames are sent at a fixed rate due to the lack of feedback through ACKs. A workaround for this, which has been commonly used, is to send unicast frames to a reference client, and put the 802.11 interfaces of the clients in monitor mode to overhear the unicast frames \cite{wevcast}. However, this requires client side device configuration changes. Also, clients in monitor mode lose their ability to send ACK for unicast frames designated for different applications. In WiCode, delivery of coded multicast is achieved by using WiCode multicast group and segment group as follows.

{\em WiCode multicast group:} In WiCode, the multicast rate of the MAC layer is set at a value lower than the maximum transmission rate at the WiFi AP, and a WiCode client can receive the multicast frames sent by WiCode-E to the group by joining the WiCode multicast group using Internet Group Management Protocol (IGMP) at the network layer, and therefore does not require device configuration changes by putting the WiFi adapter into monitor mode.

%and WiCode-E uses multicast UDP packets at the transport layer to send coded segments. A single UDP packet translates to a single multicast frame at the MAC layer. 

{\em WiCode segment group:} WiCode also creates a logical grouping of clients called segment group within the WiCode multicast group. A segment group is the group of clients for which a segment (set of UDP packets) is sent. WiCode-C of a client processes the UDP packets of a segment only if it belongs to the segment group for which the UDP packets are sent. This is illustrated in Fig. \ref{fig:multicast}. Also, the UDP packets sent to a segment group may be lost due to collisions or buffer overflow, and therefore, have to be retransmitted.

{\em Retransmission:} In WiCode, UDP packets which are lost are requested for retransmission by the clients. For the set of packets corresponding to a segment, WiCode-E waits to receive the retransmission requests of all the clients. It then attempts to reduce the number of retransmitted UDP packets by index coding those that need to be retransmitted. For example, consider a coded segment that is multicast to segment group $\{c_{i}, c_{j}, c_{k}\}.$ Now assume that $c_i$ lost packet $p_1$ and $c_j$ lost packet $p_2.$ If no other packets were lost, then WiCode-E index codes packets $p_1$ and $p_2$ and performs exactly one transmission to enable $c_i$ and $c_j$ to recover the lost packets. The general case can be reduced to a clique-cover problem of an undirected graph $G(V,E)$ \cite{indexalgorithms}, where the vertices $V = \{v_{1}, \ldots \}$ are the requested UDP packets, and an edge exists between two vertices $v_{i}$ and $v_{j}$ if the client requesting the UDP packet $v_{i}$ had received $v_{j}$ successfully and vice versa. The algorithm for coded UDP packets retransmission is illustrated in Algorithm \ref{algo:retransmit}.
\vspace{-2mm}
%UDP packets retransmission request from a client $c_{i}$ can be represented as a pair ($W(c_{i})$, $H(c_{i}))$ as described in Sec. \ref{subsec:index}, where $W(c_{i})$ is the set of UDP packets not received by $c_{i}$, and $H(c_{i})$ = $W(c_{i})^{c}$ is the set of UDP packets received by $c_{i}$. The client $c_{i}$ has to send only $W(c_{i})$.

\begin{algorithm}%[tbh]
  \caption{UDP packets retransmission}
  {\small
  \label{algo:retransmit}
  \SetAlgoLined
  %\KwResult{Write here the result }
  \While{G(V,E) has clique of size 3}{
    Find a clique $\{v_{i}, v_{j}, v_{k} \}$ of size 3\;
    XOR the UDP packets $\{v_{i}, v_{j}, v_{k} \}$ and multicast to the clients\;
    Remove $\{v_{i}, v_{j}, v_{k} \}$ from $G(V,E)$\;
  }
  Compute a maximum matching of $G(V,E)$\;
  \While{there is a pair $\{v_{i}, v_{j} \}$ in the matching}{
    XOR the UDP packets $\{v_{i}, v_{j} \}$ and multicast to the clients\;
    Remove $\{v_{i}, v_{j} \}$ from $G(V,E)$\;
  }
  Send the remaining packets from $G(V,E)$\;
  }

\vspace{-1mm}
\end{algorithm}
\vspace{-5mm}
\section{Implementation}
\label{sec:implement}

\begin{figure}[t]
\centering
\includegraphics[width=0.4\textwidth]{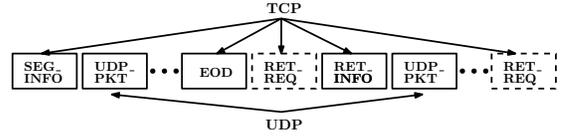}  
\vspace{-2mm}
\caption{WiCode control headers and data packets.}
\label{fig:control}
\vspace{-5mm}
\end{figure}

WiCode uses variable length control headers and data packets to deliver video segments. Fig. \ref{fig:control} shows the control headers and data packets used for coded segment delivery. WiCode uses TCP to send control headers and UDP to send segment data. To initiate a segment transmission to a segment group, WiCode-E sends $\texttt{SEG\_INFO}$ to the clients in a segment group. A segment group is identified using a 16-bit field inside $\texttt{SEG\_INFO}$. Any other information required by the clients in a segment group to decode the segments that they requested are present in $\texttt{SEG\_INFO}$. After $\texttt{SEG\_INFO}$ is sent to the clients, WiCode-E starts sending multicast UDP packets ($\texttt{UDP\_PKT}$), which is followed by $\texttt{EOD}$ to notify the end of UDP packets transmission. Each $\texttt{UDP\_PKT}$ is identified using a 16-bit $\texttt{UDP\_SEQ\_NO}$ field. Once the clients receive $\texttt{EOD}$, they send their UDP retransmission request ($\texttt{RET\_REQ}$) to WiCode-E. WiCode-E sends the information containing which UDP packets are coded together to the segment group using $\texttt{RET\_INFO}$, and then it resends the requested UDP packets. A segment transmission to a segment group is complete when all the clients in a segment group sends their retransmission request ($\texttt{RET\_REQ}$) with a field $\texttt{UDP\_PKT\_COUNT}=0$ which indicate that all the clients in a segment group have received the segment successfully. 
\vspace{-2mm}
\section{Experimental Evaluation}
\label{sec:experiments}
%\vspace{-2mm}
We now report the results of some experiments that we have performed. The aim of these experiments is to determine the gain due to coding in WiCode when coding opportunities exist between the requests of the clients. These are not meant to be a comprehensive study of achievable performance. 
\vspace{-3mm}
\subsection{Experimental setup}
\label{subsec:setup}
{\em HTTP video server:} Our HTTP video server is a PC equipped with Core i7-3770 CPU and 16GB RAM, running a python based web server. The web server hosts video files which are split into multiple segments of 4 seconds playback duration and are encoded to 5 Mbps bitrate, 1280x720 resolution, 24fps. 

{\em WiFi AP, WiCode-E:} We use TP-Link AC1750 WiFi AP, and it runs on a Linux OpenWrt firmware. The multicast frame rate of the AP is set to 24 Mbps. The WiCode-E module inside the AP is written in C++, and runs at the userspace.
%We use TP-Link AC1750 WiFi AP, and it runs on Qualcomm Atheros QCA9558 chipset with 128MB RAM, and a Linux OpenWrt firmware. The multicast frame rate of the AP is set to 24 Mbps. The WiCode-E module inside the AP is written in C++, and runs as an application at the userspace.

{\em HTTP video player, WiCode-C:} We developed a web based HTTP based video player using JavaScript. The WiCode-C module is written in C++, and runs as a Google Native Client (NaCl)\cite{native} plugin on a Chrome browser. WiCode-C uses Chrome browser's storage as segment cache.

{\em Wireless clients:} Our wireless clients are located in a University lab setting. Each client uses TP-Link WN722N WiFi adapter, and runs on Windows or Linux Ubuntu OS.
\vspace{-2mm}
\subsection{Experiment, Metrics, and Results}
\begin{figure*}%[ht]
  \centering
  \subfloat[Data transmitted]
  {\includegraphics[width=0.32\textwidth]{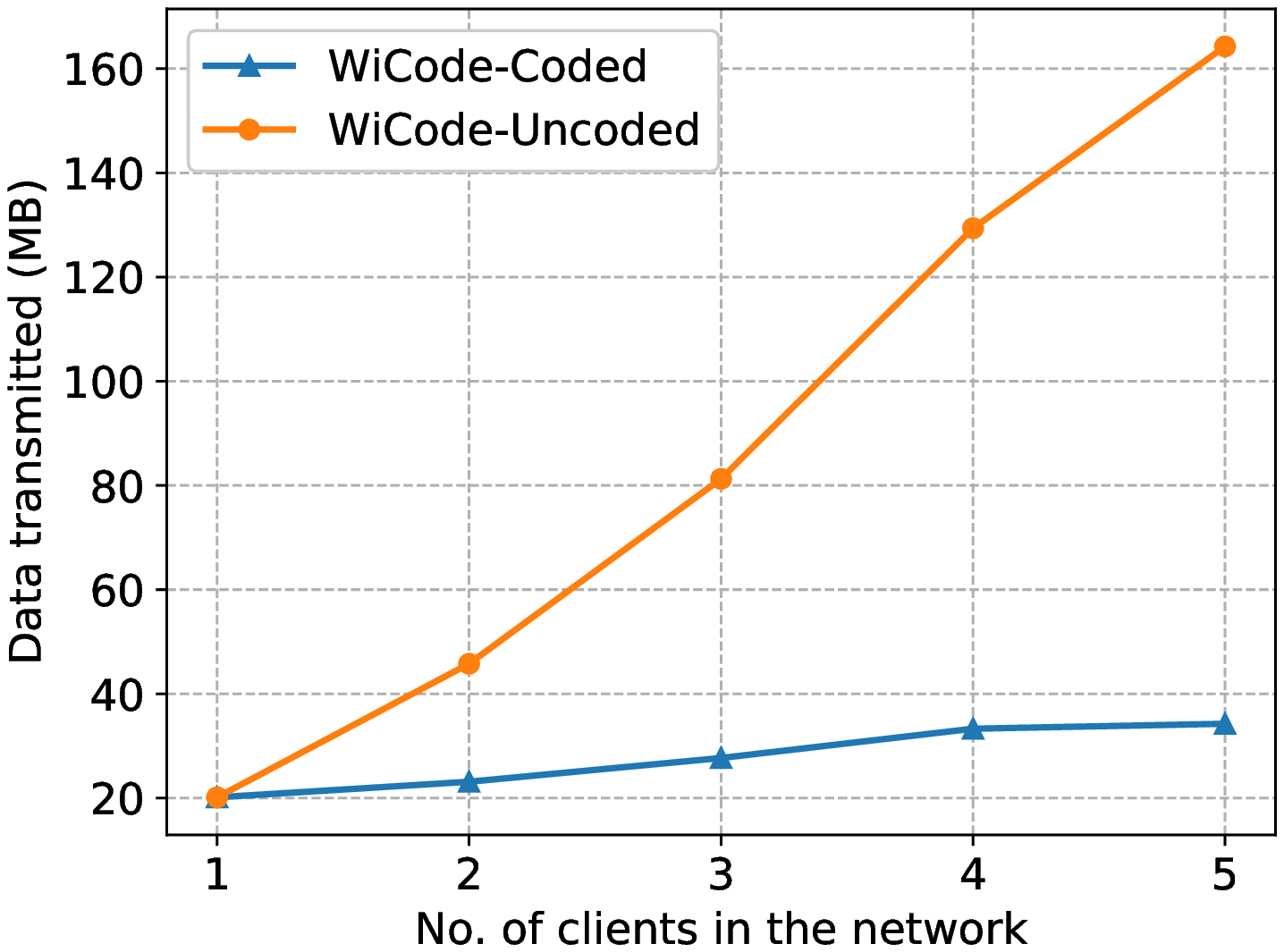}\label{fig:data_transmitted}}
  \subfloat[Coding gain]
  {\includegraphics[width=0.32\textwidth]{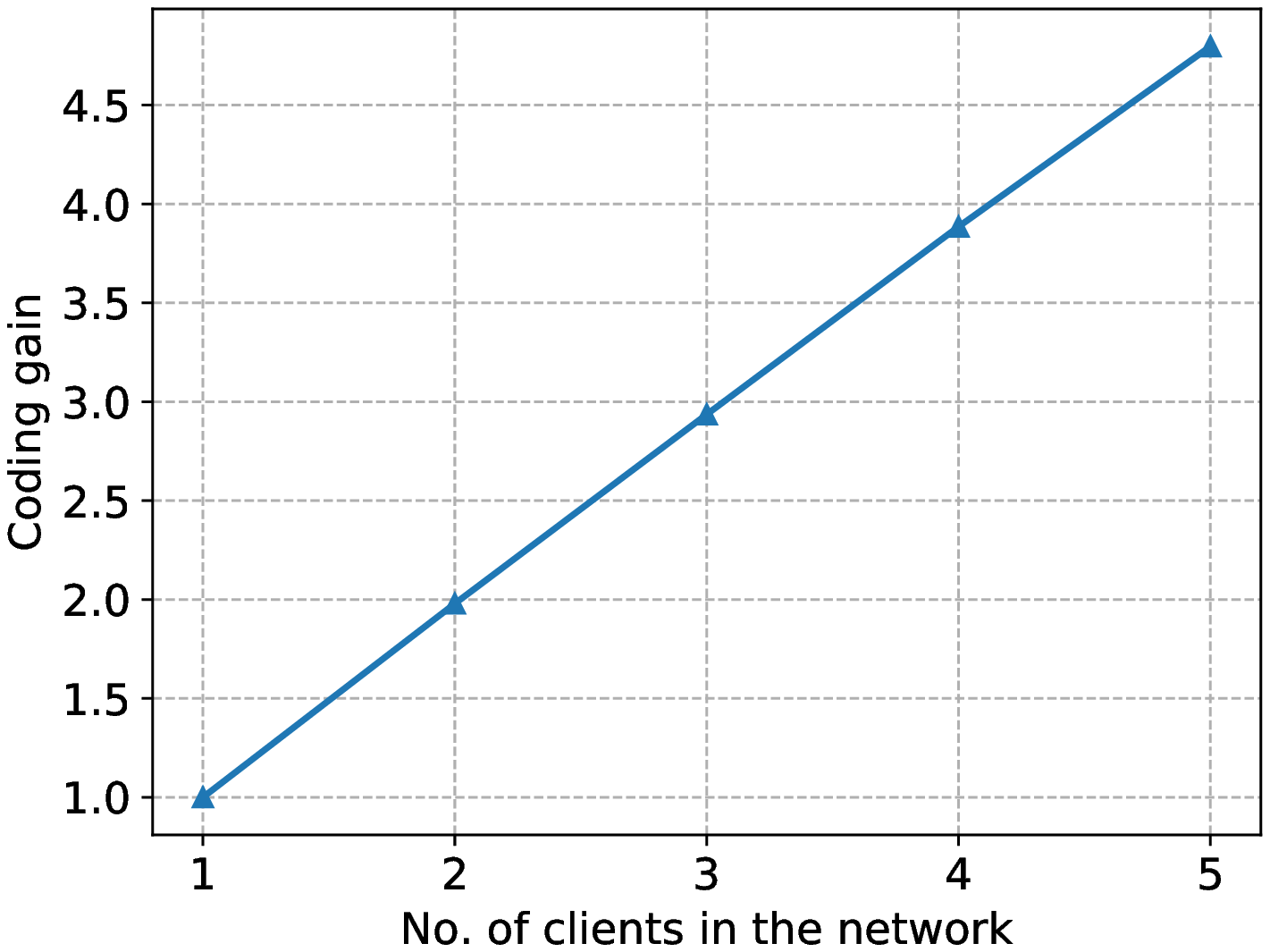}\label{fig:coding_gain}}
  \subfloat[Fraction of control bits when index coding is enabled (WiCode-Coded)]
  {\includegraphics[width=0.32\textwidth]{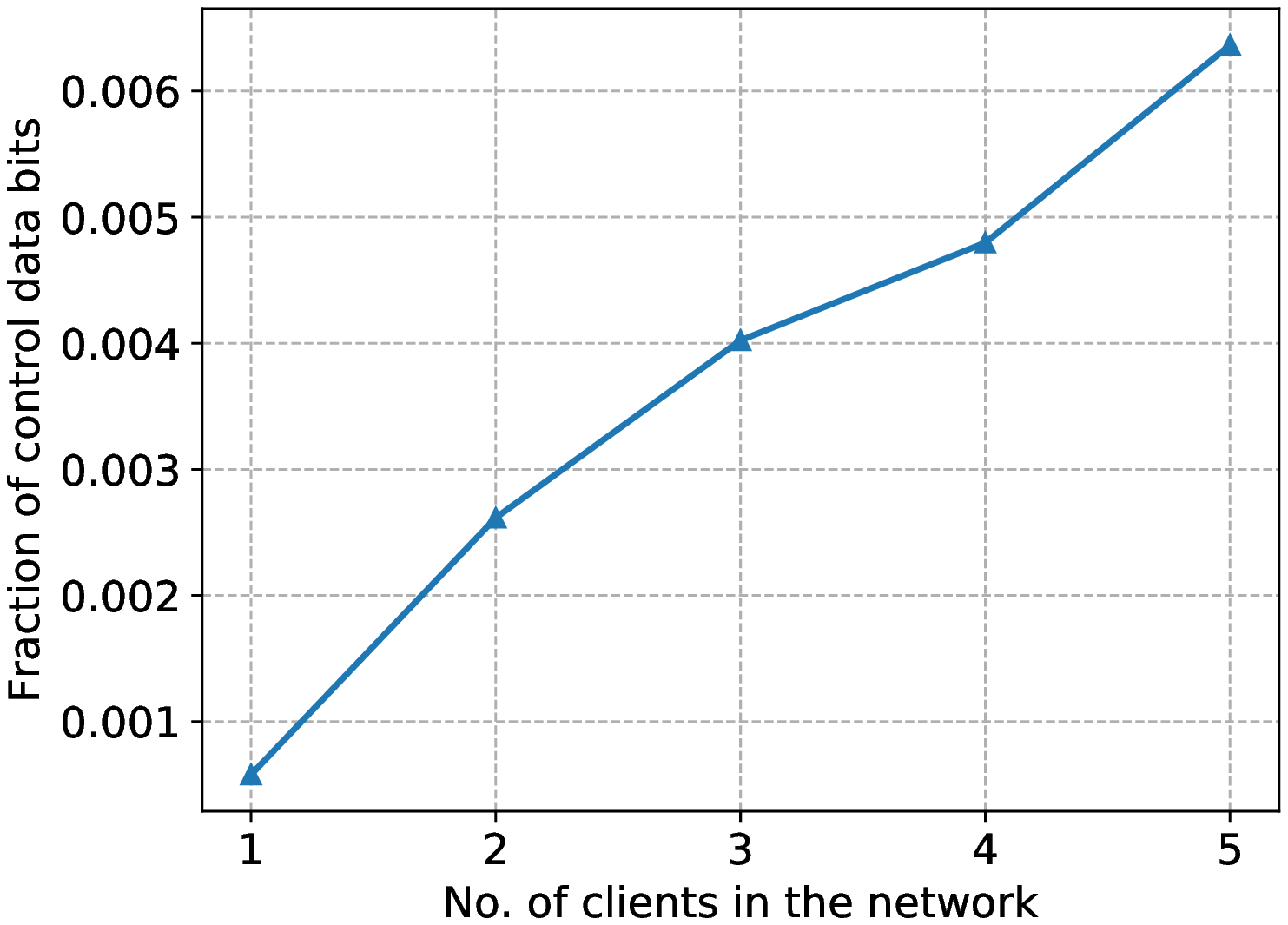}\label{fig:fraction}}
  \caption{Performance comparison of WiCode when index coding is enabled (WiCode-Coded) and disabled (WiCode-Uncoded).}
  \label{fig:fig1}
  \vspace{-7mm}
\end{figure*}

\begin{figure*}%[ht]
  \centering
  \subfloat[Throughput of HTTP, WiCode-Coded, and \newline WiCode-Uncoded ]{\includegraphics[width=0.32\textwidth]{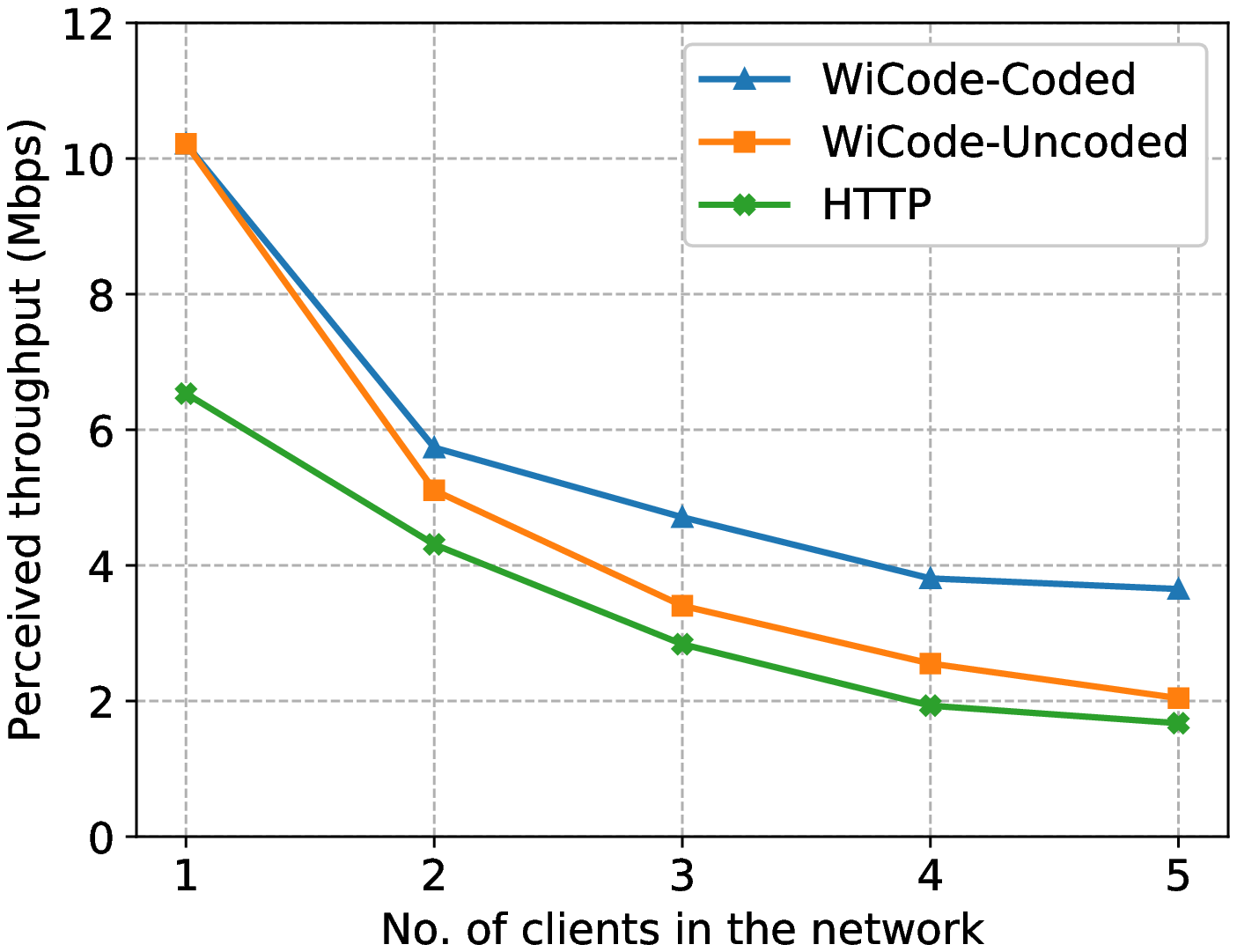}\label{fig:tr_all}}
  \subfloat[Throughput of WiCode-Coded-Variable and \newline WiCode-Coded-Fixed]{\includegraphics[width=0.32\textwidth]{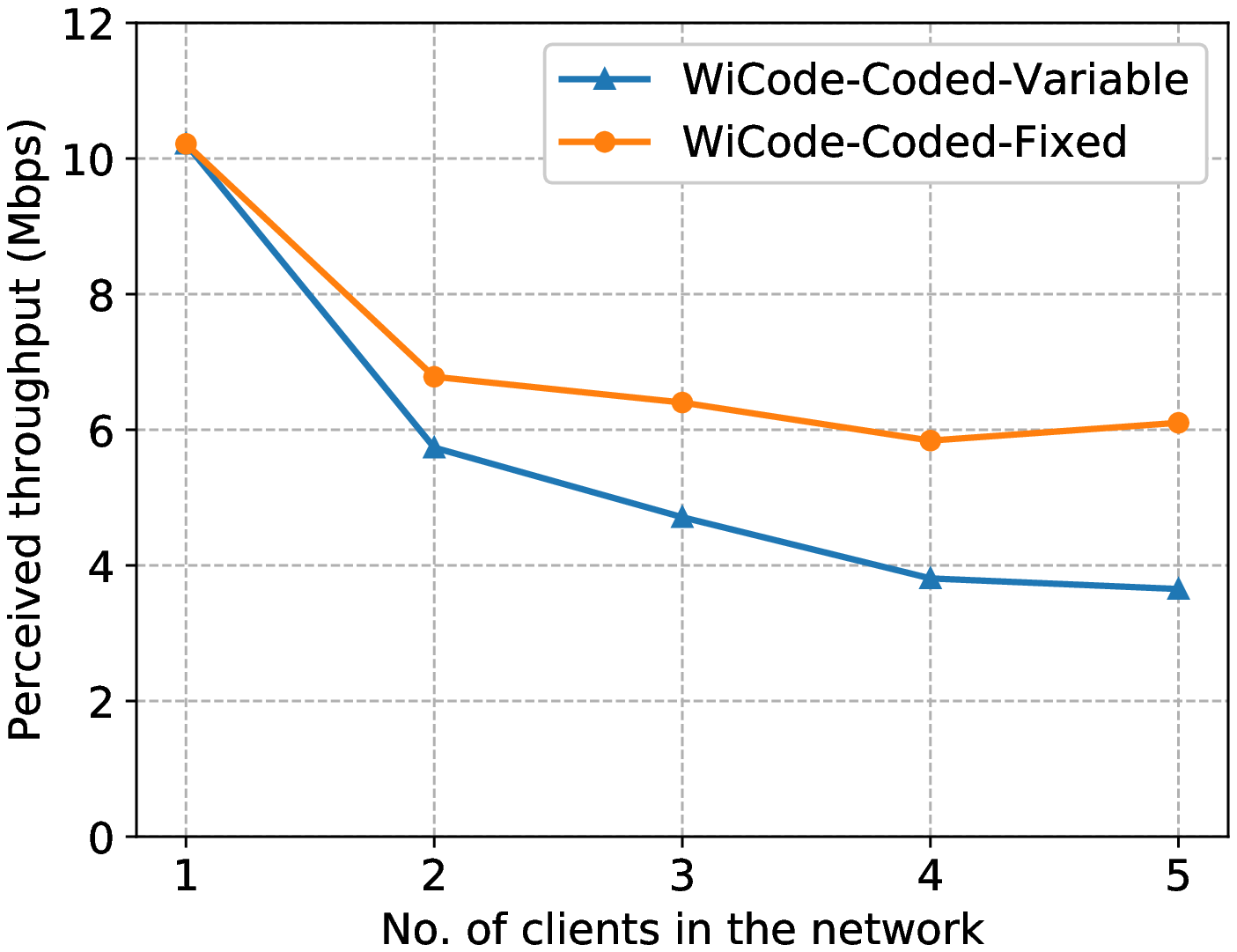}\label{fig:tr_comp}}
  \subfloat[Segment size distribution in WiCode-Coded-Variable]{\includegraphics[width=0.32\textwidth]{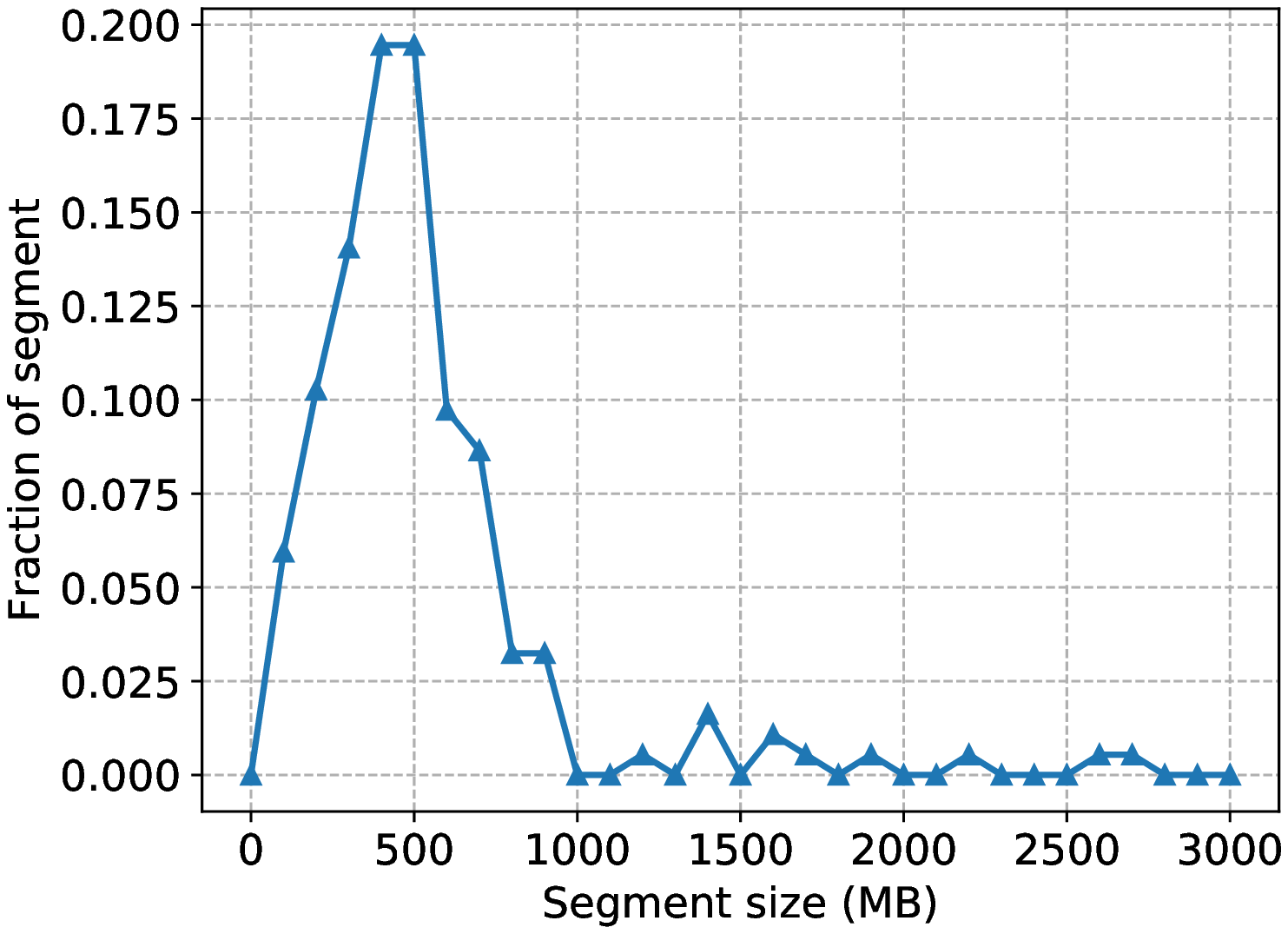}\label{fig:data_dist}}
  \caption{Throughput comparison between different scenarios in WiCode and HTTP}
  \label{fig:fig2}
  \vspace{-5mm}
\end{figure*}

%\begin{figure}
%\centering
%\includegraphics[width=0.40\textwidth]{Images/cache_read_xor.eps}    
%\caption{}
%\label{fig:cache_read}
%\end{figure}

We start with a scenario where there is only one client in the network, and the client requests video segments of a video stream. After a certain period of time, a second client is then added to the network which also starts requesting segments that can be index coded with the first client's requests. Similarly, a third client is also added whose requests can be index coded with the first two clients' requests, and this process continues up to five clients. The performance observed (when index coding is enabled) is then compared with the case when index coding is disabled. We also compare the performance of WiCode with regular HTTP streaming where clients stream videos directly from the HTTP server.

We use the following metrics to evaluate the performance of WiCode.

{\em Coding gain: } We define coding gain as the ratio of the number of transmitted data bits by WiCode-E when index coding is disabled to the number of transmitted data bits by WiCode-E when index coding is enabled, i.e.,
\begin{align*}
\text{Coding gain} &= \frac{\text{No. of TX bits with no index coding}}{\text{No. of TX bits with index coding}}
\end{align*}

Fig. \ref{fig:data_transmitted} shows the transmitted data of WiCode-Coded (index coding enabled) and WiCode-Uncoded (index coding disabled) for different number of clients in the network. We can see that in WiCode-Coded, the transmitted data remains almost the same with the increase in the number of clients as the number of segments which are coded together also increases. Whereas, in the case of WiCode-Uncoded, the data transmitted increases linearly as no coding is performed. This results in an increase in coding gain with the increase in the number of clients as shown in Fig.~\ref{fig:coding_gain}.

{\em Fraction of control bits: } We define the fraction of control bits as the ratio of the transmitted control data bits to the transmitted segment data bits. Fig. \ref{fig:fraction} shows the fraction of control data bits for different number of clients in the network when index coding is enabled. We can see that the transmitted control data is negligible compared to the transmitted segment data. However, there is a linear increase with the increase in the number of clients because the control data is unicast to all the clients in a segment group, and therefore, as the clients in a segment group increase, the transmitted control bits also increases.

{\em Average perceived throughput by the clients:} We define perceived throughput by a client as
\begin{align*}
\text{Perceived throughput} &= \frac{\text{Size of segment requested}}{\text{$T_{e}$}}
\end{align*}
where $T_e$ is the time elapsed between when a segment is requested by a client and when the segment is received and decoded successfully at the client. Fig.~\ref{fig:tr_all} shows the average perceived throughput of WiCode-Coded, WiCode-Uncoded and HTTP for different number of clients in the network. WiCode-Coded and WiCode-Uncoded are the cases when index coding is enabled and disabled respectively in WiCode, and HTTP is the case where clients stream videos directly from the HTTP server. We can see that the perceived throughput decreases with the increase in the number of clients for all the cases due to network congestion. WiCode-Coded performs better than WiCode-Uncoded and HTTP, because in WiCode-Coded multiple segments are transmitted together which reduces the network congestion. However, in WiCode-Coded, due to the asynchronous arrival of requests at WiCode-E, there are cases where requests could not be index coded which also results in a slight decline in performance.

%In the case of WiCode-UnCoded and HTTP, a client has to wait for all the other segment requests that arrived earlier at WiCode-E to be transmitted before the segment it requests can be transmitted which results in increase in $T_e$. In the case of WiCode-Coded, as a segment request is index-coded with the other requests, multiple segments are sent simultaneously in a single coded segment, the increase in $T_e$ lesser than the other cases. However, due to the asynchronous nature of arrival of requests at WiCode-E, there are cases where requests cannot be index coded which results in increase in $T_e$, and therefore also results in decline in throughput.

%\vspace{-1mm}

{\em Effect of variable length segment size on perceived throughput: } 
%As described in Sec. \ref{subsec:variable}, variable length segment size decreases the efficiency of index coding. 
The variation in segment sizes decreases the perceived throughput at the clients as a client that requests a smaller segment which is index coded with a bigger segment also has to download the entire coded segment (size of the coded segment is the size of the bigger segment), to decode the segment which results in lesser perceived throughput. Fig. \ref{fig:tr_comp} shows the perceived throughput of WiCode-Coded-Variable (variable segment size) and WiCode-Coded-Fixed (fixed segment size) for different number of clients in the network. We can see that the perceived throughput of WiCode-Coded-Fixed is better than WiCode-Coded-Variable due the absence of discrepancies in the size of segments coded together in WiCode-Coded-Fixed. Fig. \ref{fig:data_dist} shows the segment size distribution in the case of WiCode-Coded-Variable.

%WiCode-Uncoded is the case where We also show the performance of WiCode when index coding is enabled and disabled. Perceived throughput is a measure of quality of experience when a user downloads a video segment. HTTP shows the perceived throughput of a normal HTTP streaming, WiCode-Uncoded is the performance of WiCode when index coding is disabled and WiCode-Coded is the performance of WiCode when index coding is enabled. WiCode-Coded performs better than WiCode-Unocded and HTTP because in WiCode-Coded, multiple segments are coded together and send at the same time, which results in the lesser perceived time taken to receive a requested segment which results in better throughput.
\vspace{-2mm}
\section{Conclusions}
%\vspace{-2mm}
%\vspace{-1mm}
In this work, we described WiCode, a platform that improves HTTP based video content delivery over a WiFi network using coded delivery. The client side module of WiCode is implemented as a browser plugin and therefore does not require device configuration changes. WiCode significantly reduces data transmission and improves the perceived bandwidth of the clients by multicasting index coded video segments over the WiFi network. We also highlighted the design challenges and provided methods to address these challenges.

A natural question that might arise is the gain due to coding in a more general setting, where coding opportunities and cache content are generated based on the requests of the users. The gain of coding in such setting is presented in \cite{wcncarxiv}.  

\bibliographystyle{IEEEtran}
\bibliography{refer}

\begin{thebibliography}{10}
\providecommand{\url}[1]{#1}
\csname url@samestyle\endcsname
\providecommand{\newblock}{\relax}
\providecommand{\bibinfo}[2]{#2}
\providecommand{\BIBentrySTDinterwordspacing}{\spaceskip=0pt\relax}
\providecommand{\BIBentryALTinterwordstretchfactor}{4}
\providecommand{\BIBentryALTinterwordspacing}{\spaceskip=\fontdimen2\font plus
\BIBentryALTinterwordstretchfactor\fontdimen3\font minus
  \fontdimen4\font\relax}
\providecommand{\BIBforeignlanguage}[2]{{%
\expandafter\ifx\csname l@#1\endcsname\relax
\typeout{** WARNING: IEEEtran.bst: No hyphenation pattern has been}%
\typeout{** loaded for the language `#1'. Using the pattern for}%
\typeout{** the default language instead.}%
\else
\language=\csname l@#1\endcsname
\fi
#2}}
\providecommand{\BIBdecl}{\relax}
\BIBdecl

\bibitem{cisco}
\BIBentryALTinterwordspacing
Cisco, ``{Cisco visual networking index: Global mobile data traffic forecast
  update, 2015-2020},'' White Paper, 2015. [Online]. Available:
  \url{http://goo.gl/tZ6QMk}
\BIBentrySTDinterwordspacing

\bibitem{dash}
``{ISO/IEC DIS 23009-1.2, Information technology-Dynamic adaptive streaming
  over HTTP (DASH)-Part 1: Media presentation description and segment
  formats}.''

\bibitem{microsoft}
\BIBentryALTinterwordspacing
``Microsoft smooth streaming.'' [Online]. Available:
  \url{http://www.iis.net/download/smoothstreaming}
\BIBentrySTDinterwordspacing

\bibitem{adobe}
\BIBentryALTinterwordspacing
``Adobe http dynamic streaming.'' [Online]. Available:
  \url{http://help.adobe.com/en\_US/HTTPStreaming/1.0/Using/index.html}
\BIBentrySTDinterwordspacing

\bibitem{apple}
\BIBentryALTinterwordspacing
``Pantos, r., may, w. 2010. http live streaming, ietf draft (jun. 2010).''
  [Online]. Available:
  \url{http://tools.ietf.org/html/draft-pantos-http-live-streaming-04}
\BIBentrySTDinterwordspacing

\bibitem{httpdash}
S.~Thomas, ``{Dynamic Adaptive Streaming over HTTP {--:} Standards and Design
  Principles},'' in \emph{Proc. 2011 ACM MMSys}.\hskip 1em plus 0.5em minus
  0.4em\relax New York, NY, USA: ACM, 2011, pp. 133--144.

\bibitem{iscod}
Y.~Birk and T.~Kol, ``{Coding on demand by an informed source (ISCOD) for
  efficient broadcast of different supplemental data to caching clients},''
  \emph{IEEE Trans. Inf. Theory}, vol.~52, no.~6, pp. 2825--2830, June 2006.

\bibitem{indexside}
Z.~Bar-Yossef, Y.~Birk, T.~Jayram, and T.~Kol, ``{Index coding with side
  information},'' \emph{IEEE Trans. Inf. Theory}, vol.~57, no.~3, pp.
  1479--1494, March 2011.

\bibitem{fundamental}
M.~A. Maddah-Ali and U.~Niesen, ``{Fundamental limits of caching},'' \emph{IEEE
  Trans. Inf. Theory}, vol.~60, no.~5, pp. 2856--2867, May 2014.

\bibitem{decentralized}
M.~A. Maddah{-}Ali and U.~Niesen, ``{Decentralized coded caching attains
  order-optimal memory-rate tradeoff},'' \emph{IEEE/ACM Trans. Netw.}, vol.~23,
  no.~4, pp. 1029--1040, Aug 2015.

\bibitem{delaysensitive}
U.~Niesen and M.~A. Maddah-Ali, ``{Coded caching for delay-sensitive
  content},'' in \emph{Proc. 2015 IEEE ICC}, June 2015, pp. 5559--5564.

\bibitem{indexalgorithms}
M.~A.~R. Chaudhry and A.~Sprintson, ``{Efficient algorithms for index
  coding},'' in \emph{IEEE INFOCOM Workshops 2008}, April 2008, pp. 1--4.

\bibitem{wevcast}
P.~Pace and G.~Aloi, ``{WEVCast: Wireless Eavesdropping Video Casting
  Architecture to Overcome Standard MulticastTransmission in Wi-Fi Networks},''
  \emph{Telecommun. Syst.}, vol.~52, no.~4, pp. 2287--2297, Apr. 2013.

\bibitem{native}
``{Google Native Client},'' \url{https://developer.chrome.com/native-client}.

\bibitem{wcncarxiv}
L.~Chhangte, E.~Viterbo, D.~Manjunath, and N.~Karamchandani, ``Online caching
  and coding at the wifi edge: Gains and tradeoffs,'' 2020, arXiv:2001.07334.

\end{thebibliography}

\end{document}